\documentclass[twocolumn,showpacs,preprintnumbers,amsmath,amssymb]{revtex4}

\usepackage{graphicx}
\usepackage{dcolumn}
\usepackage{bm}

\begin{document}
%\begin{CJK}{UTF8}{nsung}

%\preprint{Submitted to Phys. Rev. E}

\title{Topological dynamics and dynamical scaling behavior of vortices in a two-dimensional XY model}

%\author{Wei-Kai Qi (齐维开)}
\author{Wei-Kai Qi}
\affiliation{Institute of Theoretical Physics,
Lanzhou University, Lanzhou $730000$, China}

%\author{Yong Chen (陈勇)}
\author{Yong Chen}
\altaffiliation{Email: \tt{ychen@lzu.edu.cn}}
\affiliation{Institute of Theoretical Physics,
Lanzhou University, Lanzhou $730000$, China}

\date{\today}

\begin{abstract}
By using topological current theory we study the
inner topological structure of vortices a
two-dimensional (2D) XY model and find the
topological current relating to the order
parameter field. A scalar field, $\psi$, is
introduced through the topological current
theory. By solving the scalar field, the
interaction energy of vortices in a 2D XY model
is revisited. We study the dynamical evolution of
vortices and present the branch conditions for
generating, annihilating, crossing, splitting and
merging of vortices. During the growth or
annihilation of vortices, the dynamical scaling
law of relevant length in a 2D XY model,
$\xi(t)\propto(t-t^*)^{1/z}$, is obtained in the
neighborhood of the limit point, given the
dynamic exponent $z=2$. This dynamical scaling
behavior is consistent with renormalization group
theory, numerical simulations, and experimental
results. Furthermore, it is found that during the
crossing, splitting and merging of vortices, the
dynamical scaling law of relevant length is
$\xi(t)\propto(t-t^*)$.  However, if vortices are
at rest during splitting or merging, the
dynamical scaling law of relevant length is a
constat.
\end{abstract}

% 47.32.C- Vortex dynamics
% 61.72.Cc Kinetics of defect formation and annealing
% 64.70.qj Dynamics and criticality

\maketitle
%\end{CJK}

\section{\label{sec:level1}INTRODUCTION}

Vortices play an important role in understanding
a variety of problems in physics. In the 1970s,
Kosterlitz and Thouless constructed a detailed
and complete theory of 2D systems~\cite{Kt}. The
KT phase transition theory predicts that vortices
pair unbinding will lead to a second-order
transition in 2D systems, such as superfluid
films, superconductors, and XY models~\cite{two,
Dn}. In a 2D XY model, there exist meta-stable
states corresponding to vortices, which are
closely bound in pairs below a critical
temperature. Above the critical temperature, the
paired vortices unbind and become free. Vortices
in a 2D XY model disrupt the spin alignments even
at large distances and correspond to
singularities of the order-parameter
field~\cite{Cl}.

In 2D XY model, vortices are topologically stable
configurations. It is found that the high
temperature disorder phase with an exponential
correlation is a result of the formation of
vortices. The critical temperature at which the
KT transition occurs is, in fact, that at which
vortex generation becomes thermodynamically
favorable. At temperatures below this, the system
has a power law correlation. In a low temperature
phase, vortices appear in a small density of
tightly bound dipole pairs. With an increase in
temperature, vortex pairs dissociate and become
free in the disorder phase. The evolution of
vortices plays an important role in the KT
transition of the 2D XY model.

There has been progress on the study of the
defects associated with an n-component vector
order parameter field, $\phi(\vec{r},t)$. For the
scalar case (n=1), the defects are domain walls,
which are points of the spatial dimensionality
d=1, lines for d=2, planes for d=3, etc. More
generally, for n=d, one has point defects. This
leads to vortices for n=d=2. It is interesting to
consider the appropriate form for the point
defect densities when expressed in terms of the
vector order parameter field. This has been
carried out by Halperin~\cite{Ha}, and exploited
by Liu and Mazenko~\cite{LM}; however, their
analyses are incomplete~\cite{Duan00}. In a 2D
system, a gauge field-theoretical formalism has
been developed by Kleinert~\cite{HK}.
Furthermore, the gauge theory of topological
quantum melting in a $2+1$ dimensional Bose
system was developed by Nussinov et al, and the
superfluidity and superconductivity can arise in
a strict quantum field-theoretical
setting~\cite{Z1}.

A topological field theory for topological
defects has been developed by Duan et
al~\cite{Duan}. By using the $\phi$-mapping
method and topological current theory, the
evolution of the topological defect that relates
to singularities of the order-parameter field,
such as the vortex in BEC~\cite{Duan04} and
superconductivity~\cite{Duan05}, was studied. In
this paper, we will discuss the topological
quantization and evolution of vortex in
two-dimensional XY model. We introduce a scalar
field, $\psi$, through the topological current
theory. By solving the scalar field, the
Hamiltonian of the interaction of vortices in the
2D XY model is revisited.

More recently, the dynamical behavior of the 2D
XY model following a quench of the temperature to
below the KT critical temperature has been
studied~\cite{Br01}. During the quench of a
dynamic 2D XY model, aging phenomena and dynamic
scaling behavior become extremely
important~\cite{Bo}. The assumption of dynamical
scaling for the predicted asymptotic growth law
the characteristic length, $\xi(t)\propto (t/ln
t)^{1/2}$~\cite{Br02}, which is also
characteristic of the spacing between defects.
However, scaling violations were reported.
According to an expansion in $\epsilon=4-d$ using
standard field-theoretic renormalization group
theory, it can be shown that the growing length
is given as $\xi(t)\propto t^{1/z}$ for large t,
where z is the critical exponent for equilibrium
critical dynamics~\cite{Jan}. This standard
theoretical approach also shows that the result
$\xi(t)\propto t^{1/z}$ is independent of the
initial conditions. However, this renormalization
group method does not involve the effects of
topological defects, such as vortices in the 2D
XY model.  In the recent work, it was shown that,
for the specific case of the 2D XY model,
$\xi(t)\propto (t/ln t)^{1/2}$ if free vortices
are present, while $\xi(t)\propto t^{1/2}$ if
there are no free vortices present in the initial
state~\cite{Brprl}.

Through our topological current theory for the 2D
XY model, the dynamic of vortices is studied, and
the branch conditions for generating,
annihilating, crossing, splitting and merging of
vortices are given. During the growth or
annihilation of vortices, the dynamical scaling
law of relevant length in the 2D XY model, which
is $\xi(t)\propto(t-t^*)^{1/2}$, can be obtained
in the neighborhood of the limit point. This
indicates that vortices in the 2D XY model are
the source of the scaling violations. This
dynamical scaling behavior is consistent with
numerical simulations~\cite{LD} and experimental
results~\cite{Pa}. Furthermore, we have also
found that during the crossing, splitting, and
merging of vortices, the dynamical scaling law of
relevant length in the 2D XY model, which is
$\xi(t)\propto(t-t^*)$.  It has been shown that
why the relevant length of vortices scales with
time as $t^{-1/2}$ for small values of times and
then deviate toward a linear dependence in
nematic liquid-crystal experimental
observations~\cite{Pa2}. However, if vortices are
at rest during splitting and merging, the
dynamical scaling law of relevant length is $\xi
= const$. Moreover, it is worthwhile to note that
the dynamical scaling law of vortices, which was
deduced from the topological current theory, only
depends on topological properties of the order
parameter field.

The organization of this paper is as follows. In
Sec. II, we describe the 2D XY model and the
$\phi$-mapping method, and topological current
theory is discussed. The Hamiltonian of vortices
in the 2D XY model is revisited in section III.
The evolution of vortices and dynamical scaling
discussed in section IV. Finally, in section V,
we summarize our results.

\section{\label{sec:level2}Topological Current in the 2D XY model}

The 2D XY model is a system of spins confined to
rotate in the plane of the lattice. The
Hamiltonian of the system is give as:
\begin{equation}
H_0=-J\sum_{<ij>} \textbf{S}_i
\textbf{S}_j=-J\sum_{<ij>}
\cos(\theta_i-\theta_j),
\label{eq:Hami01}
\end{equation}
where J is the strength of the nearest-neighbor
interaction. $\theta_i$ denotes the angle of the
spin on site i with respect to arbitrary polar
direction in the 2D vector space containing
spins.

By using the continuum limit, we can approximate
$\cos(\theta_i-\theta_j)$ by the first two terms
$1-(\theta_i-\theta_j)^2/2$ of the Taylor
expansion. We can express partial derivatives
through $\theta_i-\theta_j=\partial\theta$ for
the two sites i and j. This leads to the
continuum Hamiltonian:
\begin{equation}
H=H_0-E_0=\frac{J}{2}\int
d\textbf{r}(\nabla\theta)^2, \label{eq:Hami02}
\end{equation}
where $E_0=2JN$ is the energy of the completely
aligned ground state of $N$ spins.

Kosterlitz and Thouless suggested that the
disordering is caused by topological defects,
such as vortices in two-dimensional XY model,
which are characterized by a mapping from some
loop $\Gamma$ in real space onto the order
parameter space. For two-dimensional XY model,
this implies
\begin{equation}
\oint d\theta=\oint \nabla\theta d\vec{s}=2\pi W,
W=0, \pm 1,\ldots, n, \label{eq:vortex}
\end{equation}
where $W$ is the winding number of the vortex. By
introducing a 2D XY model, the local order
parameter is defined as:
\begin{equation}
\Psi=\Psi_0 e^{i\theta}=\phi^1+i\phi^2.
\label{eq:phi}
\end{equation}
Quantized vortices are topological objects
associated with topological properties of the
order parameter $\Psi$. It is worth noting that
the phase of the order parameter is undefined at
the vortex core. In other words, vortices
correspond to singularities of the
order-parameter field. We define the unit vector
field $\vec{n}$ as
\begin{equation}
n^a=\frac{\phi^a}{||\phi||},~||\phi||=\sqrt{\phi^a\phi^a},~a=1,2.
\label{eq:na}
\end{equation}
where $n^a n^a=1$. From the unit vector,
$\vec{n}(x)$, we can construct a topological
current of the order parameter field in the 2D XY
model, which carries the topological information
of $\vec{\phi}(x)$:
\begin{equation}
j^k=\frac{1}{2\pi}\epsilon^{ijk}\epsilon_{ab}\partial_{i}n^a\partial_{j}n^b,
\quad i, j, k=0, 1, 2. \label{eq:top}
\end{equation}
We will see that this current does not vanish at
the zero point of $\vec{\phi}(x)$ or the
singularities of the unit field $\vec{n}(x)$. By
using the 2D topological current theorem,
Eq.~(\ref{eq:top}) can be rewritten in the
compact form;
\begin{equation}
j^k=\delta^2(\vec{\phi})J^{k} \left(
\frac{\phi}{x} \right). \label{eq:JC}
\end{equation}
It is the important relation between the
$\delta-$like topological current of vortices and
the order parameter, $\Psi$, where
$J^{k}(\phi/x)$ is the vector Jacobian of
$\vec{\phi}$:
\begin{equation}
J^{k} \left( \frac{\phi}{x} \right) = \frac{1}{2}
\epsilon^{ijk} \epsilon_{ab} \partial_{i} \phi^a
\partial_{j} \phi^b, \label{eq:Jacobian}
\end{equation}
or
\begin{eqnarray}
J^0 \left( \frac{\phi}{x} \right) &=& \det\left(\begin{array}{cc} \partial_x\phi^1 & \partial_y\phi^1\\ \partial_x\phi^2 & \partial_y\phi^2 \end{array}\right)\nonumber,\\
J^1 \left( \frac{\phi}{x} \right) &=& \det\left(\begin{array}{cc} \partial_y\phi^1 & \partial_t\phi^1\\ \partial_y\phi^2 & \partial_t\phi^2 \end{array}\right)\nonumber,\\
J^2 \left( \frac{\phi}{x} \right) &=&
\det\left(\begin{array}{cc} \partial_t\phi^1 &
\partial_x\phi^1\\ \partial_t\phi^2 &
\partial_x\phi^2 \end{array}\right)\nonumber.
\end{eqnarray}

According to the implicit function theorem, the
Jacobian's determinant can be given as:
\begin{equation}
J^0 \left( \frac{\phi}{x} \right) = J \left(
\frac{\phi}{x} \right) \neq 0, \label{eq:Ja0}
\end{equation}
The solutions of the zero point of
$\vec{\phi}(x)$ can be generally expressed as:
\begin{equation}
x = x_l(t), \quad y = y_l(t), \quad l =
1,2,\ldots,N, \label{eq:zero}
\end{equation}
which represent $N$ zero points, $\vec{z_l}(t)$
(l=1,2,\ldots,N), or a world line of $N$ vertices
in space-time.

With the $\delta$-function theory,
$\delta^2(\phi)$, can be expanded as:
\begin{equation}
\delta^2(\phi) = \sum_{l=1}^N
\frac{\beta_l}{|J(\phi/x)_{z_l}|} \delta^2 \left(
\vec{r}-\vec{z_l}(t) \right) \label{eq:delta}
\end{equation}
where the positive integer, $\beta_l$, is called
the Hopf index of map $x\rightarrow\vec{\phi}$.
The meaning of $\beta_l$ is that when the point
$\vec{r}$ covers the neighborhood of the zero,
$\vec{z_l}$, once the vector field, $\vec{\phi}$,
covers the corresponding region for $\beta_l$
times. Using the implicit function theorem and
the definition of the vector Jacobian
(Eq.~\ref{eq:Jacobian}), we can find the velocity
of the $l$-th defect,
\begin{eqnarray}
\vec{v_l} &=& \frac{d\vec{z_l}}{dt} = \left[
\frac{\vec{J}(\phi/x)}{J(\phi/x)} \right]
_{\vec{z_l}}
\nonumber\\
\vec{J} \left( \frac{\phi}{x} \right) &=& \left[
J^1\big(\phi/x\big), J^2 \left( \phi/x \right)
\right]
\end{eqnarray}

The spatial and temporal components of the defect
current, $j^k$, can be written as the form of the
current and the density of a system of $N$
classical point particles moving in a
$(2+1)$-dimensional space-time,
\begin{equation}
\vec{j}=\sum_{l=1}^{N}\beta_l\eta_l\vec{v_l}\delta^2(\vec{r}-\vec{z_l}(t))
\label{eq:current}
\end{equation}
\begin{equation}
j^0=\rho=\sum_{l=1}^{N}\beta_l\eta_l\delta^2(\vec{r}-\vec{z_l}(t))
\label{eq:rho}
\end{equation}
where $\eta_l$ is the Brouwer degree,
\begin{equation}
\eta_l=\frac{J(\phi/x)}{|J(\phi/x)|}\bigg|_{\vec{z_l}}=\pm1.
\label{eq:eta}
\end{equation}
It can clearly be seen that
Eq.~(\ref{eq:current}) shows the movement of
vertices. The topological charge of vertices in
the 2D XY model are conserved:
\begin{equation}
\frac{\partial \rho}{\partial t}+\nabla\vec{j}=0.
\label{eq:conserved}
\end{equation}
In addition, there is a constraint of charge
neutrality:
\begin{equation}
\int\rho
d^2x=\frac{1}{2\pi}\sum_{l=1}^{N}\beta_{l}\eta_{l}=0,
\label{eq:CN}
\end{equation}
which indicates that vortices in the 2D XY model
appear in pairs.

\section{The Hamiltonian of vortices in the 2D XY model revisited}

In analogy to the velocity field in a superfluid,
the distortion field, $\vec{u}=\nabla\theta$,
carries the topological information of vortices
in the 2D XY model. By using Eq.~(\ref{eq:phi})
and (\ref{eq:na}), we can prove that
$$
\vec{u}=\epsilon_{ab}n^a\nabla n^b,
$$
and the vorticity is given as:
\begin{equation}
\nabla\times\vec{u}=\textbf{e}_i
(\epsilon^{ijk}\epsilon_{ab}\partial_{j}n^a\partial_{k}n^b)
\label{eq:vor}
\end{equation}
where $\textbf{e}_k (k=1,2,3)$ are the base
vectors in the Cartesian coordinate system.
Comparing Eq.~(\ref{eq:vor}) and
Eq.~(\ref{eq:top}), we conclude that
\begin{equation}
\nabla\times\vec{u}=2\pi j^{0}\textbf{z}.
\label{eq:vc}
\end{equation}
Therefore, in the 2D XY model, the vorticity of
distortion field, $\nabla\times\vec{u}$, can be
expressed in terms of the topological current of
the order parameter field. By comparing the
$\delta-$like topological current
Eq.~(\ref{eq:JC}) and Eq.~(\ref{eq:vc}), we have
the important relation between vorticity and the
order parameter field in the 2D XY model:
\begin{equation}
\nabla\times\vec{u}=2\pi\delta^2(\phi)J\left(\frac{\phi}{x}\right)\textbf{z},
\label{eq:VJ}
\end{equation}
From Eq.~(\ref{eq:VJ}), we see that the vorticity
of distortion field, $\nabla\times\vec{u}$, does
not vanish at the zero points of $\Psi$. The
location and direction of the $i$th vortex are
determined by the $i$th singular point,
$\vec{z_l}(t)$, and the vector Jacobian,
$J(\phi/x)$, on $\vec{z_l}(t)$, respectively. In
the absence of vorticity, there are no zero
values of the order parameter field,
$\delta^2(\phi)$ is zero, and Eq.~(\ref{eq:VJ})
becomes:
\begin{equation}
\nabla\times\vec{u}_0=0,
\label{eq:irrotationality}
\end{equation}
which is the condition of irrotationality. Thus,
Eq.~(\ref{eq:VJ}) describes both the vortex-state
and the irrotationality-state. To describe a
collection of vortices at locations
$\vec{z_l}(t)$, we substitute Eq.~(\ref{eq:rho})
into Eq.~(\ref{eq:vc}), leading to:
\begin{equation}
\nabla\times\vec{u} = 2\pi \sum_{l=1}^{N} \beta_l
\eta_l \delta^2 (\vec{r}-\vec{z_l}(t))
\textbf{z}. \label{eq:TS}
\end{equation}
It can be see that Eq.~(\ref{eq:TS}) represents N
vortices that are charged with the topological
charge, $W$. $W$ is $\beta_l\eta_l$, which
describes the inner topological structure of the
vortex. To find the solution for
Eq.~(\ref{eq:TS}), we can introduce a harmonious
scalar field, $\psi$ defining by:
\begin{eqnarray}
\nabla^2\psi&&=\partial_{i}\partial_{j}\theta-\partial_{j}\partial_{i}\theta
\nonumber\\
&&=\epsilon^{ij}\epsilon_{ab}\partial_{i}n^a\partial_{j}n^b.
\end{eqnarray}
In the absence of a vortex in the 2D XY model, we
can see that the condition,
$\partial_{i}\partial_{j}\theta=\partial_{j}\partial_{i}\theta$,
leads to $\nabla^2\psi=0$. Thus, the 2D
distortion can be written as,
$\vec{u}=\vec{u}_0-\nabla\times(\textbf{z}\psi)$,
which contains two parts: the first part,
$\vec{u}_0=\nabla\phi$, describe spin waves,
while the second part is associated with
vortices. By using 2D topological current
theorem, it can be shown that:
\begin{equation}
\nabla^2\psi=2\pi\sum_{l=1}^{N}\beta_l\eta_l\delta^2(\vec{r}-\vec{z_l}(t)).
\label{eq:SD}
\end{equation}
The scalar field, $\psi$, behaves like the
potential due to a set of charged particles. The
solution of Eq.~(\ref{eq:SD}) is given as:
\begin{equation}
\psi(x)=\sum_{l=1}\beta_l\eta_l
ln(|\vec{r}-\vec{z_l}(t)|). \label{eq:PS}
\end{equation}
The continuum Hamiltonian of the vortices in
Eq.~(\ref{eq:Hami02}) can be rewritten as:
\begin{equation}
H-\sum_{i}E_{ci}=\frac{J}{2}\int(\nabla\psi)^2d^2\textbf{x}=-\frac{J}{2}\int\psi\nabla^2\psi,
d^2\textbf{x} \label{eq:H2}
\end{equation}
where $E_{ci}$ is the core energy of the vortex.
Substituting Eq.~(\ref{eq:SD}) and (\ref{eq:PS})
into the Hamiltonian (\ref{eq:H2}), one can
obtain:
\begin{equation}
\beta H=\beta E_c\sum_{i}W_i^2-\pi
K\sum_{ij}W_iW_j ln(\vec{z_i}(t)-\vec{z_j}(t)),
\label{eq:Hami03}
\end{equation}
which is identical to the Hamiltonian of a 2D
Coulomb gas with point charges of charge
$W_i=\beta_i\eta_i$. The constraint given by
Eq.~(\ref{eq:CN}) is thus the constraint of
charge neutrality.

\section{The evolution of vortices in the 2D XY model}

The zero point of the order parameter field,
$\Psi$, which is associated with the locations of
the core of vortices, plays an important role in
describing the evolution of vortices in the 2D XY
model.  If the Jacobian determinant,
$J^0(\phi/x)\neq0$, we will have the isolated
solution of the zeros of the order parameter
field in $2+1$ dimensional space-time. But, when
$J^0(\phi/x)=0$, the above results will change in
some way and will lead to the branch process of
vortices. We denote one of the vectors Jacobians
at the zero points as $(t^*, \vec{z_l})$.
According to the values of the vector Jacobian at
the zero points of the order parameter, there are
usually two kinds of branch points, namely, the
limit points and bifurcation points~\cite{KU}.
Each kind corresponds to different cases of
branch processes.
\begin{figure}
\begin{center}
\includegraphics[width=0.4\textwidth]{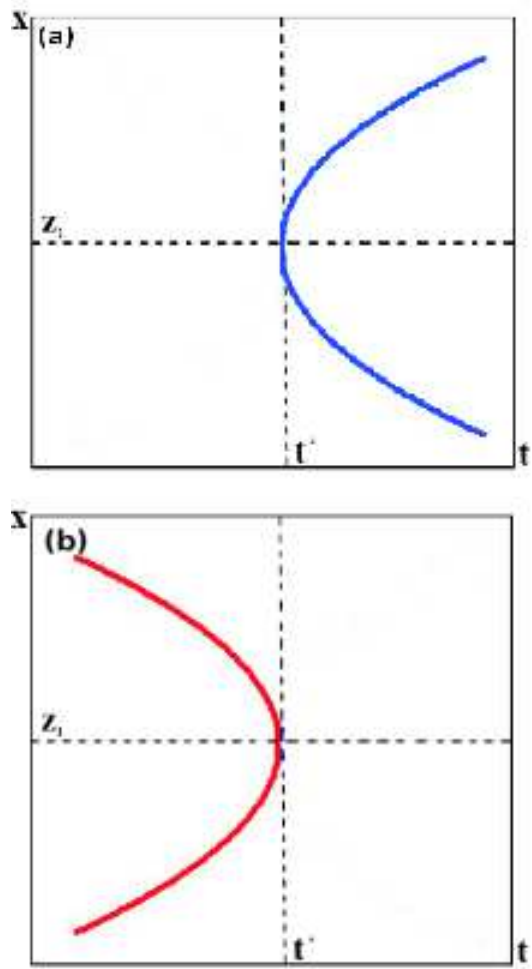}
\caption{Generating and annihilating of vortices pairs. (a) The
origin of two vortices. (b) Two vortices annihilate in collision at
the limit point. } \label{Fig:B}
\end{center}
\end{figure}

\subsection{The generation and annihilation of vortices}

We explore what will happen to vortices at the
limit point $(t^*, \vec{z_l})$. The limit points
are determined by
\begin{eqnarray}
J^{0} \left. \left(\frac{\phi}{x}\right)
\right|_{t^*, \vec{z_l}}=0, \quad J^{1}\left.
\left(\frac{\phi}{x}\right) \right|_{t^*,
\vec{z_l}}\neq0
\label{eq:condition01} \\
J^{0} \left. \left( \frac{\phi}{x} \right)
\right|_{t^*, \vec{z_l}}=0, \quad J^{2} \left.
\left(\frac{\phi}{x}\right) \right|_{t^*,
\vec{z_l}}\neq 0. \label{eq:condition02}
\end{eqnarray}
Considering the condition given by
Eq.~(\ref{eq:condition01}) and making use of the
implicit function theorem, the solution of the
zero points of $\phi(x)$ in the neighborhood of
the point ($t^*, \vec{z_l}$) is given as:
\begin{equation}
t=t(x),\quad y=y(x), \label{eq:zeropoint}
\end{equation}
where $t^*=t(z_l^1)$. In this case, one can see
that:
\begin{equation}
\left. \frac{dx}{dt} \right|_{(t^*, \vec{z_l})}=
\left. \frac{J^1(\phi/x)}{J(\phi/x)}
\right|_{(t^*, \vec{z_l})}=\infty, \label{eq:zl}
\end{equation}
or
\begin{equation}
\left. \frac{dt}{dx} \right|_{(t^*, \vec{z_l}}=0.
\label{dtdx01}
\end{equation}
The Taylor expansion of $t=t(x)$ at the limit
points $(t^*, \vec{z_l})$ is given as:
\begin{equation}
t-t^*= \left. \frac{1}{2}\frac{d^2t}{dx^2}
\right|_{t^*, \vec{z_l}}(x-z_l^1)^2,
\label{eq:MV}
\end{equation}
which is a parabola in the x-t plane. From this
equation, we can obtain two solutions $x_1(t)$
and $x_2(t)$, which give two branch solutions
(world lines of the vortices). If
$$
\left. \frac{d^2t}{dx^2} \right|_{(t^*,
\vec{z_l})}>0,
$$
we have the branch solutions for $t > t^*$ which
are related to the origin of a dipole pair.
Otherwise, we have the branch solutions for $t <
t^*$, which related to the annihilation of a
dipole pair (Fig.~\ref{Fig:B}).
\begin{figure}
\begin{center}
\includegraphics[width=0.5\textwidth]{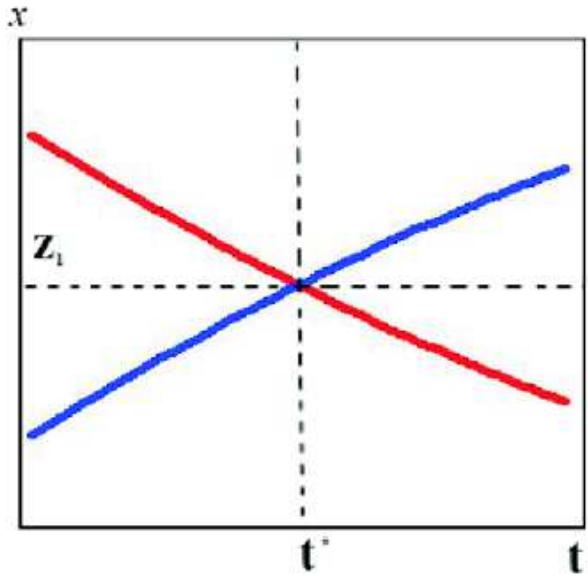}
\caption{Two vortices collide with different directions of motion at
the bifurcation point in $(2+1)$-dimensional space-time. Two world
lines intersect with different directions at the bifurcation point;
i.e., two vortices encounter at the bifurcation point.}
\label{Fig:C}
\end{center}
\end{figure}

Since the topological current is identically
conserved, the topological charges of these two
generated or annihilated vortices must be
opposite of one another at the limit points, such
as:
\begin{equation}
\beta_1\eta_1+\beta_2\eta_2=0.
\label{eq:conserved02}
\end{equation}
This indicates that the vortices always generate and annihilate in
pairs. From Eq.~(\ref{eq:MV}), one also obtains that the velocity of
the vortices is infinite when they are annihilating, which agrees
with the result introduced by Bray~\cite{Bray}. Furthermore, a new
result can be obtained that shows that the velocity of the vortices
is infinite when they are generating, which is gained only from the
topology of the order parameter field. For a limit point, it is
required that, $J^1(\phi/x)|_{t^*,\vec{z_l}}\neq0$. A bifurcation
point, on the other hand, must satisfy a more complex condition.
This case will be discussed in the following.

\subsection{Encountering, Splitting and Merging of vortices}

Now, let us turn to consider the case in which the restrictions on
the zero point $(t^*, \vec{z_l})$ are given by:
\begin{equation}
J^{k} \left. \left(\frac{\phi}{x}\right)
\right|_{(t^*, \vec{z_l})}=0, \quad k=0,1,2,
\label{eq:con}
\end{equation}
which imply an important fact that the function
relationship between t and x or y is not unique
in the neighborhood of the bifurcation point
$(t^*, \vec{z_l})$. This fact is easily seen from
\begin{equation}
\frac{dx}{dt}= \left.
\frac{J^1(\phi/x)}{J(\phi/x)} \right|_{t^*,
\vec{z_l}},~~ \frac{dy}{dt}= \left.
\frac{J^2(\phi/x)}{J(\phi/x)} \right|_{t^*,
\vec{z_l}}, \label{eq:con2}
\end{equation}
which, under Eq.~(\ref{eq:con}), directly shows the indefiniteness
of the direction of the integral curve of Eq.~(\ref{eq:con2}) at
$(t^*, \vec{z_l})$. For this reason, the point $(t^*, \vec{z_l})$ is
called a bifurcation point of the orientation order parameter.
\begin{figure}
\begin{center}
\includegraphics[width=0.5\textwidth]{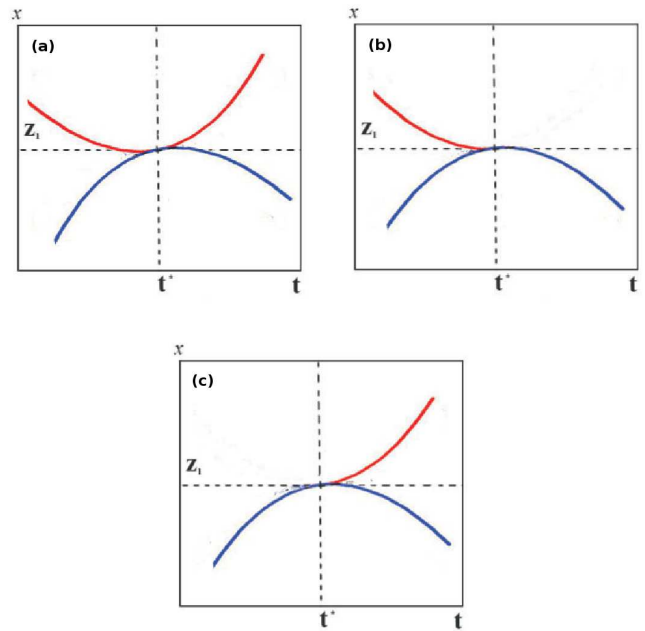}
\caption{Vortices with the same direction of motion. (a) Two world
lines tangentially contact; i.e., two vortices tangentially
encounter at the bifurcation point. (b) Two world lines merge into
one world line; i.e., two vortices merge into one vortex at the
bifurcation point. (c) One world line resolves into two world lines;
i.e., one vortex splits into two vortices at the bifurcation point.}
\label{Fig:D}
\end{center}
\end{figure}

As we know, at the bifurcation point, $(t^*, \vec{z_l})$, the rank
of the Jacobian matrix, $\left[ \partial\phi/\partial\textbf{x}
\right]$, is $1$ in the 2D vector order parameter. With the aim of
finding the different directions of all branch curves at the
bifurcation point, we assume:
\begin{equation}
\left. \frac{\partial \phi^1}{\partial y}
\right|_{t^*, \vec{z_l}}\neq0. \label{eq:bifur01}
\end{equation}
From the implicit function theorem, there is one function relation:
\begin{equation}
y=f(x,t). \label{eq:IM}
\end{equation}
According to the $\phi$-mapping theory, the Taylor expansion of the
solution of the zeros of the order parameter field in the
neighborhood of $(t^*, \vec{z_l})$ can be expressed as:~\cite{Duan}
\begin{equation}
A(x-z_l^1)^2+2B(x-z_l^1)(t-t^*)+C(t-t^*)^2+
\cdots = 0, \label{eq:two00}
\end{equation}
which leads to
\begin{equation}
A\left(\frac{dx}{dt}\right)^2+2B\frac{dx}{dt}+C=0,
\label{eq:two01}
\end{equation}
and
\begin{equation}
C\left(\frac{dt}{dx}\right)^2+2B\frac{dt}{dx}+A=0,
 \label{eq:two02}
\end{equation}
where A, B, and C are constants determined by the order parameter.
The solutions of Eq.~(\ref{eq:two01}) or Eq.~(\ref{eq:two02}) give
different directions for the branch curves (world line of the
vortices) at the bifurcation point. There are four possible cases
which demonstrate the physical meaning of the bifurcation points.
\begin{figure}
\begin{center}
\includegraphics[width=0.45\textwidth]{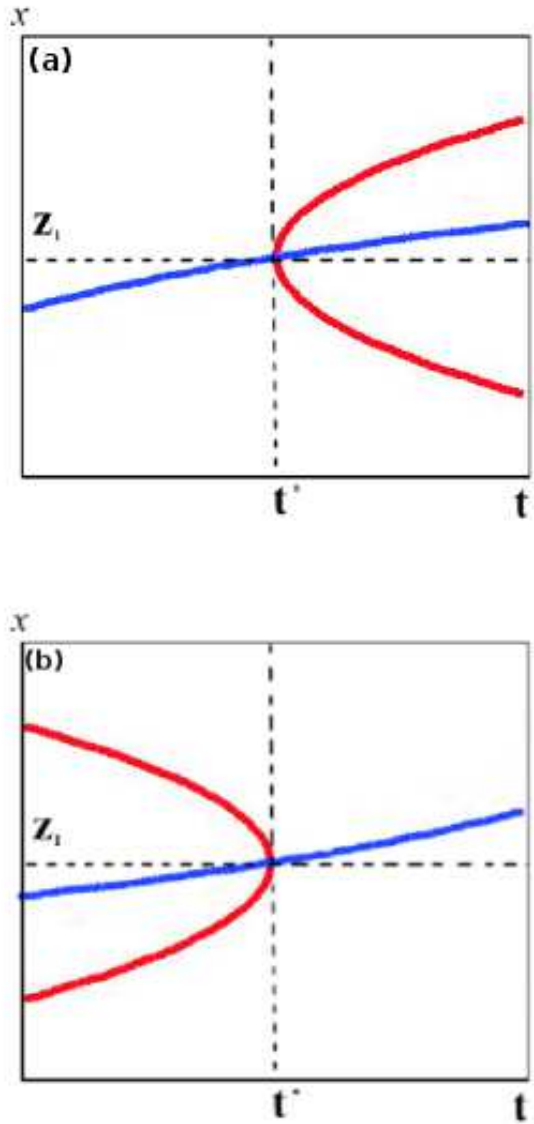}
\caption{Two important cases of Eq.~(\ref{eq:case3}). (a) One world
line resolves into three world lines; i.e., one vortex resolves into
three vortices at the bifurcation point. (b) Three world lines merge
into one world line; i.e., three vortices merge into one vortex at
the bifurcation point.} \label{Fig:E}
\end{center}
\end{figure}

Case 1 $(A\neq0)$: For $\Delta=4(B^2-AC)>0$, from
Eq.~(\ref{eq:two01}) we get two different motion directions of the
core of the vortex given as:
\begin{equation}
\left. \frac{dx}{dt}
\right|_{1,2}=\frac{-B\pm\sqrt{B^2-AC}}{A},
\label{eq:case1}
\end{equation}
which is shown in Fig. \ref{Fig:C}, where two world lines of two
vortices intersect with different directions at the bifurcation
point. This shows that two vortices encounter and then depart at the
bifurcation point.

Case 2 $(A\neq0)$: For $\Delta=4(B^2-AC)=0$, form
Eq.~(\ref{eq:two01}), we obtain only one motion direction of the
core of the vortex given as:
\begin{equation}
\left. \frac{dx}{dt} \right|_{1,2}=-\frac{B}{A},
\label{eq:case2}
\end{equation}
which includes three important cases (Fig.~\ref{Fig:D}). These three
cases are: (i) Two world lines tangentially contact; i.e., two
vortices tangentially encounter at the bifurcation point. (ii) Two
world lines merge into one world line; i.e., two vortices merge into
one vortex at the bifurcation point. (iii) One world line resolves
into two world lines; i.e., one vortex splits into two vortices at
the bifurcation point.

Case 3 $(A=0, C\neq0)$: For $\Delta=4(B^2-AC)=0$, from
Eq.(\ref{eq:two01}) we have:
\begin{equation}
\left. \frac{dt}{dx}
\right|_{1,2}=\frac{-B\pm\sqrt{B^2-AC}}{C}=0,
~-\frac{2B}{C}. \label{eq:case3}
\end{equation}
There are two important cases (Fig.~\ref{Fig:E}): (i) One world line
resolves into three world lines; i.e., one vortex splits into three
vortices at the bifurcation point. (ii) Three world line merge into
one world line; i.e., three vortices merge into one vortex at the
bifurcation point.

Case 4 (A=C=0): Equations (\ref{eq:two01}) and (\ref{eq:two02}) give
respectively:
\begin{equation}
\frac{dx}{dt}=0, \quad \frac{dt}{dx}=0.
\label{eq:case4}
\end{equation}
This case shows that two world lines intersect
normally at the bifurcation point, which is
similar to case 3. It is no surprise that both
parts of Eq.~(\ref{eq:case4}) are correct because
they give the slope coefficients of two different
curves at the same point $(t^*, \vec{z_l})$.

The remaining components of $dy/dt$ can be calculated from a
function relation of Eq.~(\ref{eq:IM}):
\begin{equation}
\frac{dy}{dt}=\frac{\partial f}{\partial
x}\frac{\partial x}{\partial t}+\frac{\partial
f}{\partial t}. \label{dydt}
\end{equation}

The above solutions reveal the evolution of the vortices. The
topological structure of the vortices is detailed in the
neighborhood of the bifurcation points of the order parameter field.
Besides the encountering of vortices, i.e., two vortices encountered
at and then depart from the bifurcation point along different branch
curves (Fig.~\ref{Fig:C} and Fig.~\ref{Fig:D}(a)), splitting and
merging of vortices are also included. When multicharged vortices
pass the bifurcation point, it may split into several vortices along
different branch curves (Fig.~\ref{Fig:D}(c), Fig.~\ref{Fig:E}(a)).
On the other hand, several vortices can merge into one vortex at the
bifurcation point (Fig.~\ref{Fig:D}(b), Fig.~\ref{Fig:E}(c)). As
before, since the topological current of the vortices is identically
conserved, the sum of topological charges of the final vortices must
be equal to that of the initial vortices at the bifurcation point,
which is given as (for fixed l):
\begin{equation}
\sum_f\beta_{lf}\eta_{lf}=\sum_i\beta_{li}\eta_{li}.
\label{eq:conserved03}
\end{equation}
This indicates that vortices with a higher value of Burgers vector
can evolve to the lower value of Burgers vector, or that vortices
with a lower value of Burgers vector can evolve to a the higher
value of Burgers vector through the bifurcation process.
Furthermore, we see that the generation, annihilation, and
bifurcation of vortices are not gradual changes, but that they start
at a critical value of a parameter, i.e., a sudden change. It is
important to note that further bifurcations are possible during the
evolution of vortices besides the cases studied in this work. It is
necessary need to assume that the terms in the Taylor series
considered above vanish and that the expansion of the field is
dominated by higher order terms.

\subsection{Dynamical scaling law of vortices in the 2D XY model}

\begin{figure}
\begin{center}
\includegraphics[width=0.5\textwidth]{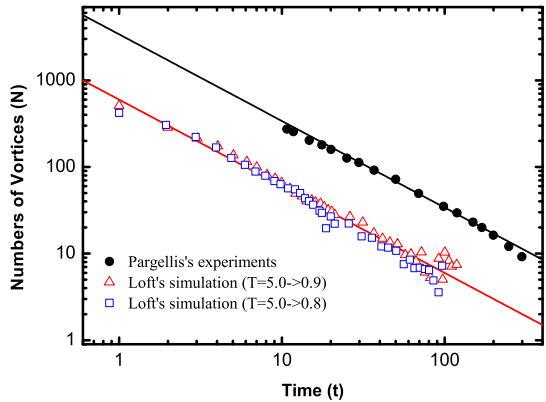}
\caption{Number of defects versus time. Experimental and simulation
results are shown. The results of simulation were the average of 20
quenches~\cite{LD}, for which $T^{*}=5.0$ initially and is suddenly
reduced to $T^{*}=0.8$ ($\square$) and $T^{*}=0.9$ ($\triangle$).
The experimental results, which are obtained by Pargellis et
al~\cite{Pa}, were averaged over nine experimental runs ($\bullet$).
Solid lines are the expected $N\propto t^{-1}$ scaling. }
\label{Fig:F}
\end{center}
\end{figure}

In the neighborhood of the limit point, we denote the scale length
$l=\nabla x$. The growth velocity or annihilation velocity of
vortices is $v=l/\nabla t$~\cite{Xu}.  From Eq.~\ref{eq:MV}), one
can obtain the scaling law as:
\begin{equation}
v\varpropto(t-t^{*})^{-1/2}. \label{scale2}
\end{equation}
It can be seen that $E_{k}\varpropto(t-t^{*})^{-1}$. In the 2D XY
model, the growth or annihilation is parameterized in terms of a
relevant characteristic length $\xi(t)$, which is also
characteristic of the mean vortex-antivortex separation distance.
From Eq.~(\ref{scale2}), we find that the relevant length, $\xi(t)$,
obeys the following:
\begin{equation}
\xi(t)\sim (t-t^{*})^{1/2}. \label{eq:scaling01}
\end{equation}
Equation (\ref{eq:scaling01}) is the dynamic scaling law of the
vortex-antivortex pairs. The low temperature equilibrium phase has
essentially no vortex pairs, and $\xi(t)$ is infinite. The
relationship between the relevant length and the metastable vortex
density below the critical temperature is given as:
\begin{equation}
\xi(t)=\sqrt{\frac{1}{\rho_v}}.
\label{eq:scaling02}
\end{equation}
then the number of vortices that satisfies the power law $N\propto
t^{-1}$. The dynamic scaling law of the vortex-antivortex pairs is
consistent with renormalization group theory and results from recent
numerical simulations using Langevin equations and Monte Carlo
methods~\cite{Brprl, LD}. The specially prepared nematic
liquid-crystal system~\cite{Pa}, which is developed, exhibits 2D XY
behavior and can be utilized to obtains the dynamical scaling
behavior of vortices (Fig.~\ref{Fig:F}). This result is consistent
with the expansion in $\epsilon=4-d$ using standard field-theoretic
renormalization group theory. This standard theory shows that the
growing length, $\xi(t)\propto t^{1/z}$ for large $t$, where $z$ is
the critical exponent for equilibrium critical dynamics~\cite{Jan}.
This standard theoretical approach also shows that the result
$\xi(t)\propto t^{1/z}$ is independent of the initial conditions.
However, this renormalization group method does not involve the
effects of topological defects, such as vortices in the 2D XY model.

From the assumptions of dynamical scaling law, it has been predicted
that the asymptotic growth law of the characteristic length is given
by $\xi(t)\propto (t/ln t)^{1/2}$~\cite{Br02}; however, scaling
violations were reported. In a recent work, it was shown that the
specific case of the 2D XY model is given as $\xi(t)\propto (t/ln
t)^{1/2}$ if free vortices are present, while $\xi(t)\propto
t^{1/2}$ if there are no free vortices present in the initial
state~\cite{Brprl}. Since the topological current is identically
conserved, there are no free vortices during the growth or
annihilation process.

In the neighborhood of the bifurcation point, we denote scale length
$\nabla x=l$. From Eqs.~(\ref{eq:case1})-(\ref{eq:case3}),  we can
then obtain the growth or annihilation velocity of the vortices,
which is given as
\begin{equation}
v \propto const. \label{eq:V1}
\end{equation}
The approximation asymptotic relation of is then
\begin{equation}
\xi(t) \propto(t-t^*). \label{eq:xi2}
\end{equation}
It can be seen that this is the reason why the relevant length of
the vortices scales with time as $t^{-1/2}$ for  small values of $t$
and then deviates towards a linear dependence in nematic
liquid-crystal experimental observations~\cite{Pa2}.

From Eq.~(\ref{eq:case4}), one can obtain
\begin{equation}
\xi(t) = const, \quad v= 0. \label{eq:xi3}
\end{equation}
It can be seen that that vortices are relatively at rest when
$\xi(t)=const$.

During the growth or annihilation of vortices, the dynamical scaling
law of the relevant length in the 2D XY model is given as
$\xi(t)\propto(t-t^*)^{1/2}$. This indicates that the vortices in
the 2D XY model are the source of the scaling violations. During the
crossing, splitting and merging of vortices, the dynamical scaling
law of relevant length in the 2D XY model is given as
$\xi(t)\propto(t-t^*)$. However, if vortices are at rest during
merging or splitting, the dynamical scaling law of relevant length
is $\xi(t) = const$. Moreover, it is worth noting that the dynamical
scaling law of vortices depends only on topological properties of
the order parameter field.

\section{Conclusions}

In summary, we have studied the inner structure and evolution of
vortices in the 2D XY model by making using of the $\phi$-mapping
topological current theory.  A scalar field, $\psi$, is introduced
through the topological current theory. By solving the scalar field,
the interaction energy of vortices in the 2D XY model is revisited.

By using the $\phi$-mapping current theory, the densities of
vortices in terms of the order parameter field in the 2D XY model
are obtained directly from the definition of the topological charges
of the vortices. The inner topological structure of the charge of
the vortices, which is characterized by the Hopf index and the
Brouwer degree, was obtained. By using the 2D topological current
theorem, the Hamiltonian of the interaction vortices in the 2D XY
model was revisited. This result is identical to the Hamiltonian of
a 2D Coulomb gas with point charges of charge $W_i=\beta_i\eta_i$.

Furthermore, we have studied the evolution of vortices in the 2D XY
model and concluded that there are crucial cases of branch processes
in the evolution of vortices when $J^0(\phi/x)=0$ and $\phi=0$,
i.e., $\eta_l$ is indefinite. It is important to note that according
to the Landau-Ginzburg approach, there is instability in the 2D XY
model at the zero point of the order parameter fields in the low
temperature phase. This indicates that vortices are unstable in the
low temperature phase. In the high temperature phase, the zero point
of the order parameter field is stable. Due to the branch condition,
the vortices generate or annihilate at the limit points and
encounter, split, or merge at the bifurcation points of the order
parameter field. This result also indicates that the velocity of the
vortices is infinite when they are being annihilated or generated,
which is obtained only from the topological properties of the order
parameter field in the 2D XY model. The scaling law of relevant
length, which is given as $R(t)\propto(t-t^*)^{1/2}$, can be
obtained in the neighborhood of the bifurcation point during the
growth or annihilation of vortices. During the crossing, splitting
and merging of vortices, the dynamical scaling law of relevant
length is $\xi(t)\propto(t-t^*)$.  However, if vortices are at rest
during merging or splitting, the dynamical scaling law of relevant
length is $\xi = const$. The dynamical scaling law of vortices only
depends on topological properties of the order parameter field.

\appendix

\section{Topological Current theory\label{appendixA}}

We study a two-component vector order parameter,
$\vec{\phi}=(\phi^1,\phi^2)$, over the base manifold, M (in this
paper $M=R^2\bigotimes R$):
\begin{equation}
\Psi=\phi^1+i\phi^2, \quad \phi^a=\phi^a(x,y,t),
\quad a=1,2. \label{eq:psi}
\end{equation}
Quantized defects are topological objects associated with
topological properties of the order parameter $\Psi$. It is worth
noting that the phase of the order parameter is undefined at the
vortex core. In other words, vortices correspond to singularities of
the order-parameter field. Let us define the unit vector field,
$\vec{n}$, as:
\begin{equation}
n^a=\frac{\phi^a}{||\phi||}, \quad
||\phi||=\sqrt{\phi^a\phi^a}, \quad a=1,2.
\label{eq:Ana}
\end{equation}
where, $n^a n^a=1$. From the unit vector, $\vec{n}(x)$, we can
construct a topological current of the order parameter field in the
2D XY model, which carries the topological information of
$\vec{\phi}(x)$:
\begin{equation}
j^k=\frac{1}{2\pi}\epsilon^{ijk}\epsilon_{ab}\partial_{i}n^a\partial_{j}n^b,
\quad i, j, k=0, 1, 2. \label{eq:Atop}
\end{equation}
It can be seen that this current does not vanish only at the zero
point of $\vec{\phi}(x)$, or the singularities of the unit field
$\vec{n}(x)$. Obviously, the current given by Eq.(\ref{eq:Atop}) is
identically conserved:
\begin{equation}
\partial_k j^{k}=0.
\label{eq:aconserved}
\end{equation}

Suppose there is a vortex located at $z_i$. The topological charge
of the defect is defined by the Gauss map, $n:
\partial\sum_i\rightarrow S^1$:
\begin{equation}
W(\phi,
z_i)=\frac{1}{2\pi}\oint_{\partial\sum_i}\epsilon_{ab}n^a
dn^b. \label{eq:winding}
\end{equation}
Using Stokes' theorem in the exterior differential form, one can
deduce that:
\begin{equation}
W(\phi,z_i)=\frac{1}{2\pi}\oint_{\sum_i}\epsilon_{ab}\epsilon^{ij}\partial_{i}n^a
\partial_{j}n^b d^2x.
\label{eq:winding2}
\end{equation}

\section{Two-dimensional topological current Theorem\label{appendixB}}

By using the $\phi$-mapping method, we can prove that the
topological current given by Eq.~(\ref{eq:Atop}) is a $\delta$-like
current:
\begin{equation}
j^{k}=\delta(\vec{\phi})J^{k}\left(\frac{\phi}{x}\right).
\label{eq:jk}
\end{equation}
From the above formula, it can be seen that the topological current
does not vanish only at the zero points of the vector field
$\vec{\phi}(x)$.

Substituting Eq.~(\ref{eq:Ana}) into Eq.~(\ref{eq:Atop}) and
considering that:
\begin{equation}
\partial_{k}n^{a}=\frac{\partial_k\phi^{a}}{||\phi||}-\phi^a\partial_{k}\left(\frac{1}{||\phi||}\right).
\label{eq:pkna}
\end{equation}
we have
\begin{equation}
j^{k}=\frac{1}{2\pi}\epsilon^{ijk}\epsilon_{ab}\partial_{i}\phi^{a}\partial_{j}^{b}\frac{\partial}{\partial\phi^a}\frac{\partial}{\partial\phi^b}(\ln(||\phi||)),
\label{eq:jk2}
\end{equation}
If we define the Jacobian as:
\begin{equation}
\epsilon^{ab}J^{k}\left(\frac{\phi}{x}\right)=\epsilon^{ijk}\partial^{j}\phi^{a}\partial^{k}\phi^{b},
\label{eq:Jac}
\end{equation}
and, by virtue of the Laplacian relation in $\phi$ space~\cite{GS}
\begin{equation}
\triangle_{\phi} \left( \ln(||\phi||) \right) =
2\pi\delta^2(\vec{\phi}), \label{eq:Lapla}
\end{equation}
where
\begin{equation}
\triangle_{\phi}=\frac{\partial}{\partial\phi^a}\frac{\partial}{\partial\phi^a}
\nonumber
\end{equation}
is the two-dimensional Laplacian operator in $\phi$ space, we obtain
a $\delta$-function like current.  This current is given by:
\begin{equation}
j^{k} = \delta (\vec{\phi}) J^{k} \left(
\frac{\phi}{x} \right). \label{eq:jk3}
\end{equation}
From the above formula, one can see that the topological current
does not vanish only at the zero points of the vector field
$\vec{\phi}(x)$. Therefore, it is essential to investigate the
solutions of $\vec{\phi}(x)=0$.

\begin{acknowledgments}
We would like to thank Y. X. Liu, T. Zhu, and Y. S. Duan for their
helpful discussion. Y.C. was supported by the SRF for ROCS, SEM, and
by the Fundamental Research Fund for Physics and Mathematics of
Lanzhou University.
\end{acknowledgments}

\end{document}